\documentclass[
 reprint,
 superscriptaddress,
 amsmath,
 amssymb,
 aps,
 prb
]{revtex4-2}

\usepackage[colorlinks=true,allcolors=blue]{hyperref}
\usepackage{graphicx}
\usepackage{dcolumn}
\usepackage{bm}
\usepackage{esint}
\usepackage{mathtools}

\usepackage[dvipsnames]{xcolor}

\newcommand{\order}[3]{#1_{#3}^{(#2)}}

\begin{document}

\title{Magnetoplasmon-Mediated Resonant Photogalvanic Effect \\ in a Gated Strip of 2D Electrons}

\author{D.A. Rodionov}%
\email{rodionov.da@phystech.edu}
\affiliation{Kotelnikov Institute of Radioengineering and Electronics of Russian Academy of Sciences, Moscow, 125009 Russia}

\author{S.G. Timchenko}%
\affiliation{Kotelnikov Institute of Radioengineering and Electronics of Russian Academy of Sciences, Moscow, 125009 Russia}
\affiliation{National Research University Higher School of Economics, Moscow, 109028 Russia}

\author{I.V. Zagorodnev}%
\affiliation{Kotelnikov Institute of Radioengineering and Electronics of Russian Academy of Sciences, Moscow, 125009 Russia}
\affiliation{National Research University Higher School of Economics, Moscow, 109028 Russia}

\date{\today}

\begin{abstract}
We theoretically investigate a nonlinear response to a linearly polarized monochromatic electromagnetic wave incident at an angle on a two-dimensional (2D) electronic system (ES) in the form of an infinite strip. The 2D ES is situated on a dielectric substrate near a perfectly conducting metal electrode (gate). The entire system is subjected to an external perpendicular constant magnetic field. We use Maxwell's equations for electromagnetic waves, while the electrons are described within the hydrodynamic approximation using Euler's equations and neglecting electromagnetic retardation effects. The incident electromagnetic wave excites magnetoplasmons in the strip. The fully screened limit is considered when all characteristic dimensions of the system, including the plasmon wavelengths, are much larger than the distance to the gate. This limit allows the linear response to be determined fully analytically. 
Due to the nonlinear hydrodynamic (convective) term, the excited magnetoplasmon oscillations give rise to a DC current along the strip and a voltage across it. Surprisingly, the relationship between the photocurrent and the photovoltage in resonance is exactly the same as in the classical Hall effect. The photovoltage is a monotonic function of the magnetic field. However, the photocurrent exhibits a minimum, which occurs at specific wavevector directions.
\end{abstract}

\maketitle

\section{Introduction}

The photogalvanic effect is a fundamental phenomenon that notably forms the basis for the operation of solar cells, various photodetectors, and other optoelectronic devices \cite{Safa2013,Eginligil2024}. It involves the generation of a DC electric current or voltage in a system when exposed to incident light.

One actively investigated possibility for its realization is the resonant photogalvanic effect based on plasma excitations in 2D electron systems, such as metal–oxide–semiconductor field-effect transistors, first proposed more than twenty years ago \cite{Dyakonov2002}. The idea is to compress incident radiation into confined two-dimensional plasmons propagating in the field effect transistor channel and to rectify them, for example, by exploiting the natural (convective) hydrodynamic nonlinearity \cite{Dyakonov1993}. The greatest interest in this arises for practical applications in the terahertz frequency range. 

Experimentally, the resonant photoresponse has been detected in almost all high-mobility 2D electron systems, including graphene \cite{Knap2002subterahertz,Knap2002terahertz,Peralta2002,Knap2004,Otsuji2004,Bandurin2018,Soltani2020}. In all these devices, the 2D plasmon frequency was tuned to the radiation frequency by changing its concentration via a voltage applied to the metallic electrode (gate). However, from the very first works, it was shown that the plasmon frequency can also be governed by a magnetic field perpendicular to the 2D system \cite{Theis1977,Muravev2015,Muravev2025}. Moreover, the electromagnetic power absorbed by a gated 2D ES in cyclotron resonance increases quadratically with the magnetic field \cite{Zabolotnykh2021}. If the response grows in the linear regime, one may also expect an increase in the nonlinear response.

Theoretical studies of the resonant photogalvanic effect, starting from the earliest works, have primarily relied on effectively one-dimensional models, where the external electric field induces purely one-dimensional electron motion in the transistor channel \cite{Dyakonov1993,Dyakonov2002,Rozhansky2015,Svintsov2018,Nikulin2021}. In practice, however, one often encounters substantially two-dimensional systems where the theory of plasma oscillations is strongly complicated, even in the case of the simplest description of 2D ESs in terms of the local Ohm law and the Drude conductivity. In the present study, we consider a two-dimensional electron gas confined within an infinite strip, where electron motion is possible both longitudinally (along the strip) and transversely (across it). The key advantage of this geometry is that, in the fully screened limit, i.e., when all characteristic lengths of the system are much greater than the distance to the gate, it allows for the exact description of linear plasma oscillations, even in the presence of a magnetic field \cite{Rodionov2023,Rodionov2024}. Therefore, it allows us to obtain well-controlled approximations for determining the nonlinear response of the system (quadratic in the exciting field).

\section{Key equations}

\begin{figure}
    \centering
    \includegraphics[width=\linewidth]{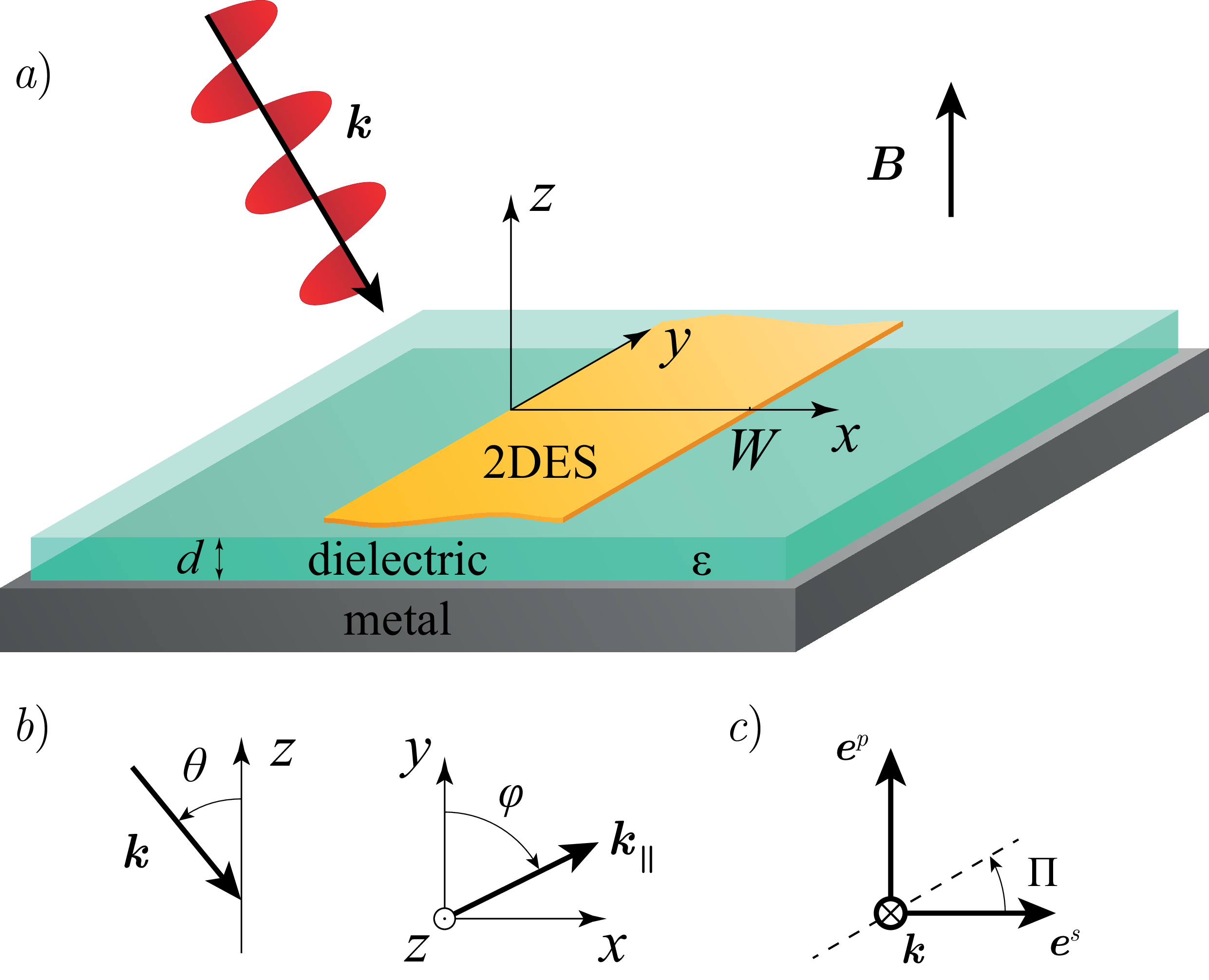}
    \caption{Schematic view of the system under consideration. a) The 2D ES is located in the plane $z = 0$ between the vacuum and the substrate  with dielectric permittivity $\varepsilon$ and the perfectly conducting metal gate on the backside. The system is placed in a magnetic field $\bm{B}$ perpendicular to the 2D ES plane. b) The orientation of the wavevector is defined the incident $\theta$ and azimuthal $\varphi$ angles. c) The angle $\Pi$ describes the polarization of the incident wave in the basis of the $s$- and $p$-polarization unit vectors. The electric field of the incident wave oscillates along the dashed line.}
    \label{fig:1}
\end{figure}

We study the response of the charge carriers in a 2D ES, which has the form of an infinite strip with width $W$. This strip is separated from a perfectly conducting metallic back gate by a dielectric substrate of permittivity $\varepsilon \ge 1$. The system is schematically represented in Fig.~\ref{fig:1}a.

By the external electromagnetic field, we understand a wave that exists in the system in the absence of the 2D ES. In our case, this field is the superposition of the incident electromagnetic wave and the wave scattered by the substrate and the metal. As the incident wave, we consider a plane monochromatic wave oscillating with frequency $\omega$ and propagating with wavevector $\bm{k}$:
\begin{equation}
    \bm{k} = \frac{\omega}{c} \left(\sin\varphi \sin\theta, \cos\varphi \sin\theta, -\cos\theta\right)^T.
    \label{eq:wavevector}
\end{equation}
Here $\theta \in [0,\pi/2]$ is the angle of incidence, and the azimuthal angle $\varphi \in [0,2\pi)$ specifies the orientation of the tangential (in‑plane) component of the wavevector relative to the strip edge (see Fig.~\ref{fig:1}b). The electric field of the incident wave is:
\begin{gather}
    \bm{\mathcal{E}}^{inc} (\bm{r},z,t) = \frac{1}{2} E_{0} \left(\cos\Pi \, \bm{e}^s + \sin\Pi \, \bm{e}^p\right) \, e^{i \bm{k}_{||}\cdot\bm{r} + i k_{z} z - i\omega t} + c.c.,
    \nonumber
    \\
    \bm{e}^{s} = \left(\cos\varphi, -\sin\varphi, 0\right)^T,
    \nonumber
    \\
    \bm{e}^{p} = \left(\sin\varphi \cos\theta, \cos\varphi \cos\theta, \sin\theta\right)^T.
    \nonumber
    \label{eq:E_inc}
\end{gather}
Here we introduce the polarization angle $\Pi \in [0,\pi)$ and the vectors $\bm{e}^s$ and $\bm{e}^p$, which correspond to the $s$- and $p$-polarization unit vectors (orts), respectively, as shown in Fig.~\ref{fig:1}c. The vectors $\bm{e}^s$ and $\bm{e}^p$ are perpendicular to the wavevector $\bm{k}$.
The reflected radiation can be determined using the complex amplitude reflection coefficients, $r_s$ and $r_p$ for the $s$ and $p$-polarized waves, respectively. We calculate them in App.~\ref{app}. In the fully screened limit, they are equal to:
\begin{equation}
    r_s = -1 - 2i k_z d, \quad r_p = 1 + \frac{2i}{\varepsilon}\left[k_z + (\varepsilon-1)\frac{\omega^2}{k_z c^2}\right]d
    \label{eq:r_coefficients}
\end{equation}
with the condition of applicability
\begin{equation}
    \cos\theta \gg \frac{\varepsilon - 1}{\varepsilon}\frac{\omega d}{c}.
    \label{eq:r_coefficients_cond}
\end{equation}
In deriving these coefficients, we assumed that the $s$‑polarization unit vector of the reflected wave coincides with that of the incident wave, while the $p$-polarization unit vector of the reflected wave is defined as $(-\bm{e}_{||}^{p}, e_z^p)$. We take into account only the tangential component of the electric field in the 2D ES plane, as only this component governs the dynamics of charge carriers:
\begin{multline}
    \bm{E}^{ext} (\bm{r},t) = \frac{1}{2}E_0\left[(1 + r_s)\cos\Pi\,\bm{e}_{||}^{s} + \right.
    \\
    \left.(1 - r_p)\sin\Pi\,\bm{e}_{||}^{p}\right]e^{i k_y y - i\omega t} + c.c.
    \label{eq:electric_field_ext}
\end{multline}

Let us turn to discussing the basic equations that fully describe the dynamics of charge carriers in the considered system. We denote the charge and current densities in the 2D ES by $\rho$ and $\bm{j}$, which are
\begin{equation}
    \rho(\bm{r},t) = e\left(n_s  + \delta n\right(\bm{r},t)),\quad \bm{j}(\bm{r},t) = \rho(\bm{r},t) \bm{v}(\bm{r},t),
    \label{eq:charge_and_current}
\end{equation}
Here $\bm{v}$ is the local two-dimensional velocity of the charge carriers, $n_s$ and $\delta n$ are the stationary two-dimensional unperturbed concentration and its deviation, respectively. The current and charge densities satisfy Euler equations:
\begin{gather}
    \begin{multlined}
        m\frac{\partial}{\partial t}\bm{v}(\bm{r},t) + m \left(\bm{v}(\bm{r},t)\cdot\nabla\right)\bm{v}(\bm{r},t) + m \gamma \bm{v}(\bm{r},t) =
        \\
        e \bm{E}(\bm{r},t) + \frac{e}{c}B_z \hat R \bm{v}(\bm{r},t),
    \end{multlined}
    \nonumber
    \\
    \frac{\partial}{\partial t}\rho(\bm{r},t) + \nabla\cdot \bm{j}(\bm{r},t) = 0,
    \label{eq:Euler_equation}
\end{gather}
where the right-hand side of the second equation is the Lorentz force acting on the charge carrier with the effective mass $m$, the charge $e$, and the electron relaxation rate $\gamma$. The matrix $\hat{R}=i\sigma_y$, where $\sigma_y$ is the Pauli $y$ matrix. In the standard manner, we neglect the compressibility of the electron fluid by omitting the pressure gradient term \cite{Fetter1986,Gunyaga2023}.

Additionally, we use the fact that the total number of electrons is conserved in the system and that no lateral current flows in or out, which implies that the current at the edges of the strip is zero. Thus, the boundary conditions are
\begin{equation}
    j_x(\bm{r},t)\Big|_{x=0;W} = 0,
    \quad
    \int\limits_{0}^{W}\delta n(\bm{r},t) dx = 0.
    \label{eq:origin_boundary_conditions}
\end{equation}

We discuss the quasistatic regime when the wavelength of the exciting field $2\pi c/\omega$ exceeds the characteristic length of the 2D ES, i.e., the width $W$. In this regime, one can neglect the magnetic field induced by the dynamics of charge carriers in the 2DES, as well as the magnetic component of the external electromagnetic waves. Therefore, the magnetic field is taken to be a constant, homogeneous perpendicular component $B_z$.

The electric field is the superposition of the external and induced fields. The latter can be found from Maxwell's equation \cite{Jin2016}. We work in the fully screened limit, where all characteristic length scales of the system — such as the strip width and plasmon wavelength — are much smaller than the substrate thickness $d$. In this limit, the connection between the induced electric field and the charge density is local \cite{Chaplik1972,Fetter1986,Rodionov2025} and is given by:
\begin{equation}
    \bm{E}(\bm{r},t) = \bm{E}^{ext}(\bm{r},t) - \frac{4\pi d}{\varepsilon} \nabla\rho(\bm{r},t).
    \label{eq:E_tot}
\end{equation}


Notice, Eqs.~\eqref{eq:charge_and_current},~\eqref{eq:Euler_equation},~\eqref{eq:E_tot} with the boundary conditions \eqref{eq:origin_boundary_conditions} can be reduced to a system of equations for the deviation of the concentration and the local velocity of charge carriers:

The system of Eqs.~\eqref{eq:charge_and_current},~\eqref{eq:Euler_equation}, ~\eqref{eq:E_tot} can be reduced to equations for the deviation in carrier density and the local carrier velocity:
\begin{gather}
    \begin{cases}
        \frac{\partial}{\partial t}\bm{v}(\bm{r},t) + \gamma \bm{v}(\bm{r},t) + v_{pl}^2 \nabla \delta\tilde{n}(\bm{r},t) - s\omega_c \hat R \bm{v}(\bm{r},t) +
        \\
        \hfill
        \left(\bm{v}(\bm{r},t)\cdot\nabla\right)\bm{v}(\bm{r},t) =\frac{e}{m}\bm{E}^{ext}(\bm{r},t),
        \\
        \frac{\partial}{\partial t}\delta\tilde{n}(\bm{r},t) + \nabla\cdot \left(1 + \delta\tilde{n}(\bm{r},t)\right)\bm{v}(\bm{r},t) = 0,
    \end{cases}
    \label{eq:system}
\end{gather}
where $v_{pl}^2 = 4\pi n_s e^2 d/(\varepsilon m)$ is the plasmon group velocity, $\omega_c = |e B_z|/(mc)$ is the cyclotron frequency, $s = sgn(e B_z)$, and $\delta\tilde{n} = \delta n/n_s$ is the dimensionless deviation of the concentration. 

The boundary condition \eqref{eq:boundary_condition} requires that $v_x(\bm{r},t) \left(1 + \delta \tilde{n}(\bm{r},t) \right)$ is equal to zero at the edges $x = 0, W$. This condition can be expressed as four distinct constraints on the velocity and the density deviation. However, since we are only interested in the regime of weak density perturbations, i.e., $|\delta \tilde n(\bm{r},t)| \ll 1$, only one variant remains possible: the normal component of the velocity must vanish at the boundaries. As a result, the boundary conditions simplify to:
\begin{equation}
    v_x(\bm{r},t) \Big|_{x = 0; W} = 0, 
    \quad
    \int\limits_{0}^{W}\delta \tilde n(\bm{r},t) dx = 0.
    \label{eq:boundary_condition}
\end{equation}

The system of Eqs.~\eqref{eq:system} and conditions \eqref{eq:boundary_condition} represents the key result of this section and forms the basis for the solutions to be developed in the subsequent sections.

\section{Linear response}

We will seek solutions in the form of a power series expansion with respect to the external electric field. Quantities linear in the field are denoted with a superscript (1), while those quadratic in the field are denoted with a superscript (2). Consequently, any quantity $a$ -- such as velocity, concentration deviation, and so on -- can be expressed as follows:
\begin{gather}
     a(\bm{r},t) = \frac{1}{2}\order{a}{1}{}(x) e^{i k_y y - i \omega t} + c.c. + \order{a}{2}{}(x) + ...,
    \nonumber
\end{gather}
where the notation $c.c.$ denotes the complex conjugate, $\order{a}{1}{}(x)$ is a complex-valued function, and $\order{a}{2}{}(x)$ is real. The linear response is characterized by terms oscillating at the driving frequency $\omega$, while the quadratic response comprises both a component at the second harmonic $2\omega$ and a time-independent, non-oscillatory contribution. Since we are only interested in the DC response, we retain only the constant (time-independent) term at the second order in the external field.

Let us discuss the linear response. Collecting first order terms in the incident amplitude from system \eqref{eq:system} yields:
\begin{gather}
    \begin{cases}
        \hat\sigma^{-1}(\omega)\order{\bm{v}}{1}{}(x) + v_{pl}^2 \nabla_1 \order{\delta\tilde{n}}{1}{}(x) = \frac{e}{m} \bm{E}^{ext}(x),
        \\
        -i\omega \order{\delta\tilde{n}}{1}{}(x) + \nabla_1 \cdot \order{\bm{v}}{1}{}(x) = 0.
    \end{cases}
    \label{eq:system_linear}
\end{gather}
Here, $\nabla_{n} = \left(\partial/\partial x, i n q_y\right)^T$, while $\hat\sigma(\omega)$ is the dynamical conductivity normalized by $n_s e^2/m$. Within the framework of the dynamical Drude model, it is given by the following expression:
\begin{equation}
    \hat\sigma(\omega) = \frac{\left(\gamma - i\omega\right)\hat I + s\omega_c \hat R}{\left(\gamma - i\omega\right)^2 + \omega_c^2},
\end{equation}
where $\hat I$ is the 2x2 identity matrix. Expressing $\delta\tilde n$ from the second equation of system \eqref{eq:system_linear} and substituting it into the first, we obtain the following operator equation for the velocity:
\begin{gather}
    \left(\hat\Omega + i \gamma \hat I\right)\order{\bm{v}}{1}{}(x) = i\frac{e}{m} \bm{E}^{ext}(x),
    \\
    \hat\Omega = \omega \hat I + \frac{v_{pl}^2}{\omega} \nabla_1 \otimes \nabla_1 - i s \omega_c \hat R,
\end{gather}
where $\otimes$ is the tensor product. Although the obtained equation can be solved exactly using ordinary differential equation methods, we propose an alternative approach that yields physically intuitive approximations, which are useful for generalizations to second (and higher) orders. The operator $\hat\Omega$ is Hermitian in the Hilbert space equipped with the scalar product
\begin{equation}
    \langle \bm{f}|\bm{g}\rangle = \int\limits_{0}^{W}\overline{\bm{f}}(x) \cdot \bm{g}(x) \frac{dx}{W},
\end{equation}
where the normal components of vector functions $\bm{f}(x)$ and $\bm{g}(x)$ vanish at the boundaries $x = 0$ and $x = W$. Therefore, its eigenfunctions form an orthogonal basis suitable for expanding the first-order velocity field $\order{\bm{v}}{1}{}$. Moreover, the eigenfunctions can be expressed as the gradient of a scalar function, which in turn satisfies a Sturm–Liouville problem \cite{Rodionov2025}.

This requires solving the eigenvalue equation $\hat\Omega\bm{f} = \Omega\bm{f}$ and finding the eigenvalues $\Omega$ and eigenfunctions $\bm{f}$ satisfying the boundary condition $f_x(0) = f_x(W) = 0$. 

Solving this problem yields eigenvalues that are indexed by an integer 
$n$ and take the following form:
\begin{equation}
    \Omega_{\pm |n|} = 
    \begin{cases}
        \omega - \frac{v_{pl}^2 k_y^2}{\omega}, & n = 0,
        \\
        \omega - \frac{v_{pl}^2 k_n^2}{2\omega} \mp \sqrt{\left(\frac{v_{pl}^2 k_n^2}{2\omega}\right)^2 + \omega_c^2}, & n \ne 0,
    \end{cases} 
    \label{eq:eigenvalues}
\end{equation}
Here, we introduce the wave vector magnitude $k_n = \sqrt{(n\pi/W)^2 + k_y^2}$, where the $x$-component of the wave vector is quantized by the width of the strip. The normalized eigenfunction for a given integer $n$ can be written as
\begin{gather}
    \bm{f}_n(x) = \left[(\omega - \Omega_n) + i s\omega_c \hat{R}\right]\nabla_{1}\psi_n(x),
    \label{eq:eigenfunctions}
    \\
    \psi_n(x) =
    \begin{cases}
        \left(\frac{v_{pl}^2 k_y}{\omega \omega_c W}\sinh \frac{\omega \omega_c W}{v_{pl}^2 k_y}\right)^{-\frac{1}{2}} e^{s\frac{\omega_c \omega}{v^2 k_y} x}, & n = 0,
        \\
        \frac{\sqrt{2}\left[\frac{n \pi}{W}(\omega-\Omega_n) \cos \frac{n\pi}{W} x + s\omega_c k_y \sin\frac{n\pi}{W} x\right]}{k_n\sqrt{\left(\left(\omega - \Omega_n\right)^2 + \omega_c^2\right)\left(\left(\frac{n\pi}{W}\right)^2\left(\omega - \Omega_n\right)^2 + k_y^2 \omega_c^2\right)}}
        , & n \ne 0.
    \end{cases}
    \nonumber
\end{gather}

\begin{figure}
    \centering
    \includegraphics[width=\linewidth]{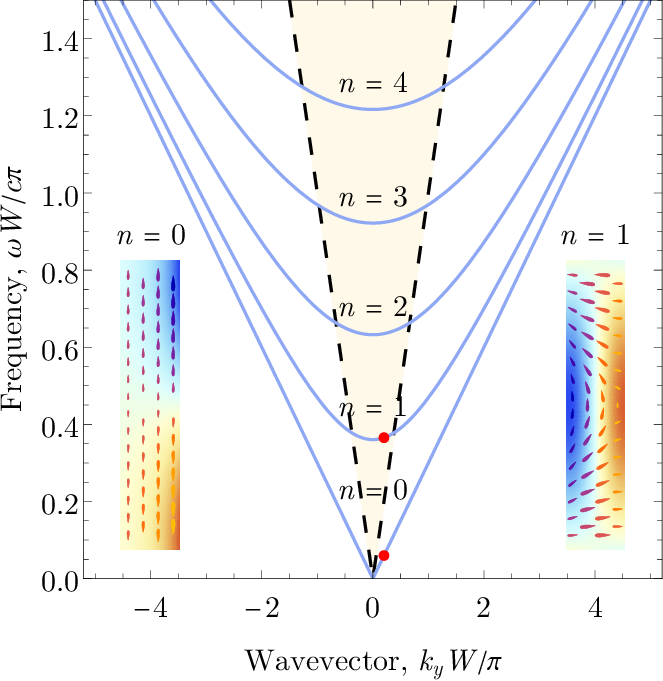}
    \caption{Dispersion of plasmons~\eqref{eq:dispersion} in the strip in the fully screened limit (solid lines) at $v_{pl} = 0.3c$, $k_y=0.2\pi/W$ and $\omega_c = 0.2cW/\pi$. The dashed lines corresponds to the light cone $c|k_y|$ inside which (colored area) plasmon excitation is possible by the plane wave. Insets present the charge (color map) and current (drops) distributions for the edge magnetoplasmon $n = 0$ and the first ``bulk'' plasmon $n = 1$ at the strip length equaling a half longitude wavelength $\pi/k_y$. }
    \label{fig:2}
\end{figure}

Consequently, the velocity is expressed as a series:
\begin{equation}
    \order{\bm{v}}{1}{}(x) = i\frac{e}{m} \sum\limits_{n = -\infty}^{\infty} \frac{\langle \bm{f}_n|\bm{E}^{ext}\rangle}{\Omega_n + i\gamma} \bm{f}_n(x)  \equiv \sum\limits_{n = -\infty}^{\infty} \bm{v}_n (x).
    \label{eq:series}
\end{equation}
Let us discuss the physical meaning of the eigenfunctions and eigenvalues of the operator $\hat\Omega$. Evidently, each term in \eqref{eq:series} exhibits a pole as a function of frequency, corresponding to the resonant velocity response under plasmon excitation. For the sake of clarity, let us consider only positive frequencies $\omega>0$. In the absence of damping ($\gamma=0$), the poles define the resonance frequencies at which the eigenvalues vanish, and these frequencies correspond to the plasmon dispersion relations:
\begin{equation}
    \omega_{0}(k_y) = v_{pl}|k_y|, \quad \omega_{n \ge 1}(k_y) = \sqrt{v_{pl}^2 k_n^2 + \omega_c^2}.
    \label{eq:dispersion}
\end{equation}
The plasmon with $n = 0$ is the edge magnetoplasmon. Its charge and current density are localized near the edge of the strip. The plane wave can only excite ``bulk'' plasmon modes ($n \ge 1$) as the edge magnetoplasmon dispersion lies below the light cone. Fig.~\ref{fig:2} shows the plasmon dispersion relation along with a schematic representation of the corresponding charge and current distributions. Near any plasmon frequency,  the resonant denominator in the series \eqref{eq:series} is proportional to $\omega - \omega_n(k_y) + i (\gamma/\Omega_n ')$, where $\Omega_n '$ is the derivative of the eigenvalue $\Omega_n$ with respect to the frequency at the plasmon frequency $\omega_n(k_y)$. It follows that the halfwidth of the plasmon resonance is
\begin{equation}
    \Delta\omega_{n} = \frac{\gamma}{\Omega_n'} = \frac{\gamma}{2}\left(1 + \frac{\omega_c^2}{v_{pl}^2 k_n^2 + \omega_c^2}\right).
\end{equation}
For bulk plasmons, the value changes monotonically from $\gamma/2$ in the weak magnetic field to $\gamma$ at the cyclotron resonance \cite{Volkov2016,Zagorodnev2023}. Thus, the eigenvalues $\Omega_n$ contain complete information about the frequency and damping rate of the magnetoplasmons.

\section{DC Photocurrent and Photovoltage}

We now calculate the DC response of the 2D ES. Collecting terms proportional to the second order in the amplitude of the incident wave, we have
\begin{equation}
    \begin{cases}
        \hat\sigma^{-1}(0)\order{\bm{v}}{2}{}(x) + v_{pl}^2 \nabla_0 \order{\delta\tilde{n}}{2}{}(x) = 
        \\
        \hfill
        -\frac{1}{2} \text{Re}\left[\left(\order{\overline{\bm{v}}}{1}{}(x) \cdot \nabla_1\right) \order{\bm{v}}{1}{}(x)\right],
        \\
        \nabla_0 \cdot \left(\order{\bm{v}}{2}{}(x) + \frac{1}{2} \text{Re}\left[\order{\delta\tilde n}{1}{} \order{\overline{\bm{v}}}{1}{}(x)\right]\right) = 0, 
    \end{cases}
    \label{eq:conductivity}
\end{equation}
From the second equation -- which corresponds to the continuity equation for the non-oscillating part of the current -- we determine $\order{v}{2}{x}(x)$. Using this, we find $\order{v}{2}{y}$ and $\partial\order{\delta\tilde n}{2}{}/\partial x$, and then  determine the current density and the electric field using \eqref{eq:charge_and_current} and \eqref{eq:E_tot}. The longitudinal DC electric field (along the strip) and the transverse current are zero. In order to make the notation shorter and physically clearer, we introduce two distinct contributions to the DC current: a ``hydrodynamic'' contribution, which originates from the convective derivative in the Euler equation~\eqref{eq:Euler_equation}, and a second contribution not related to the convective derivative but arising from the contribution to the current due to the product of the linear corrections to velocity and concentration (``non‑hydrodynamic''). We denote them by the superscript $hd$ and $nhd$, respectively. As a result, we obtain the following expressions:
\begin{gather}
    \order{j}{2}{x} = 0, 
    \qquad
    \order{j}{2}{y} = j^{nhd} + j^{hd},
    \nonumber
    \\
    j^{nhd} = \frac{1}{2} e n \text{Re}\left[\order{\delta\tilde n}{1}{}\left( \order{\overline{v}}{1}{y} + \order{\overline{v}}{1}{x} \frac{\sigma_{xy}(0)}{\sigma_{xx}(0)}\right)\right],
    \nonumber
    \\
    j^{hd} = -\frac{1}{2} \frac{e n \sigma_{xx}(0)}{\sigma_{xx}^2(0) + \sigma_{xy}^2(0)} \text{Re}\left[\left(\order{\overline{\bm{v}}}{1}{} \cdot \nabla_1\right) \order{v}{1}{y}\right],
    \label{eq:current_quad}
\end{gather}
and, similarly, for the electric field:
\begin{gather}
    \order{E}{2}{x} = E^{nhd} + E^{hd},
    \qquad
    \order{E}{2}{y} = 0,
    \nonumber
    \\
    E^{nhd} = -\frac{1}{2} \frac{m}{e\sigma_{xx}(0)} \text{Re}\left[\order{\delta\tilde n}{1}{} \order{\overline{v}}{1}{x}\right],
    \nonumber
    \\
    E^{hd} = \frac{1}{2} \frac{m}{e} \text{Re}\left[\left(\order{\overline{\bm{v}}}{1}{} \cdot \nabla_1\right) \left(\order{v}{1}{x} + \order{v}{1}{y} \frac{\sigma_{xy}(0)}{\sigma_{xx}(0)}\right)\right].
    \label{eq:field_quad}
\end{gather}
For brevity, we omit the $x$-dependence in Eqs.~\eqref{eq:current_quad} and \eqref{eq:field_quad}. Due to the perpendicular orientation of the current and the electric field, the quadratic response results in the absence of their contribution to Joule heating.

Of practical interest are the integral current along the strip $J$ and the transverse voltage $U$ which are:
\begin{equation}
    J = \int\limits_{0}^{W}\order{j}{2}{y}(x)dx, 
    \quad 
    U =  -\int\limits_{0}^{W}\order{E}{2}{x}(x)dx.
\end{equation}
To estimate them in the vicinity of the $n$-th plasmon resonance, where they are most significant, we can further assume that $\order{\bm{v}}{1}{}(x) \approx \bm{v}_n(x)$. In the case, the current $J^{hd}$ and $U^{nhd}$ tuned out to be equal to zero and, correspondingly, $J^{(2)} = J^{nhd}$ and $U^{(2)} = U^{hd}$. Since the linear concentration deviation and the velocity have the plasmon poles (according to the definition $\bm{v}_n(x)$ in Eq.~\eqref{eq:series}), the current and voltage exhibit similar resonance behavior.

We write out their peak values divided by the average intensity of the incident wave power $I_0 = c |E_0|^2 / (8\pi)$ in the following form:
\begin{gather}
    \frac{J_n}{I_0} = \frac{eW}{m c \gamma} \frac{\order{A}{1}{n}}{1 + \frac{\omega_c^2}{\omega_n^2}} \cos\varphi \sin\theta \cos\theta,
    \label{eq:photocurrent_common}
    \\
    \frac{U_n}{I_0} = s\frac{\omega_c W}{|e| n_s c \gamma} \frac{\order{A}{1}{n}}{1 + \frac{\omega_c^2}{\omega_n^2}} \cos\varphi \sin\theta \cos\theta.
    \label{eq:photovoltage_common}
\end{gather}
Here $\order{A}{1}{n}$ is the absorption coefficient of the linear response at the resonance frequency $\omega_n(k_y)$ and equals a ratio of the average joule-heating power per the strip length to the average radiant power of the incident wave per the strip length $I_0 W \cos\theta$, i.e., in our case:
\begin{equation}
    \order{A}{1}{n} = \frac{4\pi}{\cos\theta} \frac{n_s e^2}{m \gamma c} \left|\frac{\langle \bm{f}_n | \bm{E}^{ext}\rangle}{E_0}\right|_{\omega = \omega_n(k_y)}^2
    \label{eq:A_coefficient_common}
\end{equation}

It follows from the expressions obtained that the photocurrent is proportional to the component of the wavevector along the strip and therefore vanishes at normal incidence. This effect, known as the photon‑drag effect, is associated with the transfer of wave momentum to charge carriers \cite{Ivchenko2005}. According to Eqs.~\eqref{eq:photocurrent_common} and~\eqref{eq:photovoltage_common}, the ratio of the photovoltage to the photocurrent equals the Hall resistivity for any $x$‑dependent spatial profile of the external incident electric field:
\begin{equation}
    \frac{U_n}{J_n} = s\frac{m\omega_c}{e^2 n_s} = \frac{B_z}{e n_s c}.
\end{equation}

Up to this point, in obtaining the expressions above, we did not take into account the explicit form of the wave vector \eqref{eq:wavevector}, assuming it to be arbitrary. We now make a final simplification, dictated by the quasistatic limit, which is associated with the smallness of the strip width W compared to the wavelength of the exciting field. In this limit, we can set $e^{ik_x x} \approx 1$, which essentially corresponds to a uniform electric field along the $x$ coordinate. Since the expressions \eqref{eq:current_quad} and \eqref{eq:field_quad} are already proportional to $k_y$, we put $k_y = 0$ in $\omega_n(k_y)$ and $\bm{f}_n$. This implies that, as defined in Eq. \eqref{eq:eigenfunctions}, two components of any eigenfunction are even when $n$ is odd (and odd when $n$ is even). Therefore, the projection $\langle \bm{f}_n | \bm{E}^{ext} \rangle$ is different from zero only if the number $n$ is odd. Therefore, in what follows we consider only even modes and, for brevity, introduce the notation $\omega_n \equiv \omega_n(k_y = 0)$. As a result, we obtain the following expression for the absorption coefficient:
\begin{gather}
    \order{A}{1}{n} = \tilde{A} \frac{\sin^2\left(\varphi - \varphi_0 \right) + \frac{\omega_c^2}{v_{pl}^2}\left(\frac{W}{n\pi}\right)^2}{1 + \frac{\omega_c^2}{\omega_n^2}},
    \nonumber
    \\
    \tilde{A} = \frac{128\pi}{\cos\theta}\frac{n_s e^2}{m \gamma c}\frac{v_{pl}^2}{c^2}\frac{d^2}{W^2}\left(\frac{\cos\theta \cos\Pi}{\sin\varphi_0}\right)^2.
    \label{eq:A_coefficient}
\end{gather}
Here, the angle $\varphi_0 \in [-\pi/2,\pi/2]$ describes the rotation of a straight line , along which the tangential component of the external electric field oscillates, from the tangential component of the wave vector $\bm{k}_{||}$ and is given by the following expression:
\begin{equation}
	\varphi_0 = -\arctan\left(\frac{\varepsilon \cos\theta \cot\Pi}{\varepsilon - \sin^2\theta}\right).
\end{equation}

Let us discuss the obtained expression \eqref{eq:photocurrent_common} together with Eq.~\eqref{eq:A_coefficient}. At the weak magnetic field, i.e., satisfying condition $\omega_c \ll v_{pl}(n\pi/W)$, we have expansion
\begin{multline}
    \frac{J_{n}}{I_0}\Bigg|_{\omega_c \ll v_{pl}\left(\frac{n\pi}{W}\right)} = \frac{eW}{m c \gamma} \frac{\tilde{A}}{2} \cos\varphi \sin2\theta \Bigg[\sin^2\left(\varphi - \varphi_0 \right) +
    \\
     \frac{\omega_c^2}{v_{pl}^2}\left(\frac{W}{n\pi}\right)^2 \cos 2(\varphi - \varphi_0) + ...\Bigg]
\end{multline}
Note that the direction of the wave vector significantly determines the photocurrent behavior. From this expression we can see that in the absence of a magnetic field the photocurrent does not depend on the mode number $n$ and is different from zero only if $\sin(\varphi - \varphi_0) \neq 0$. The quadratic magnetic addition can either increase or decrease the photocurrent. It is positive when $\cos 2(\varphi - \varphi_0) > 0$ and negative when $\cos 2(\varphi - \varphi_0) < 0$. In the classic strong magnetic field such as $\omega_c \gg v_{pl}(n\pi/W)$ the photocurrent is
\begin{equation}
    \frac{J_{n}}{I_0}\Bigg|_{\omega_c \gg v_{pl}\left(\frac{n\pi}{W}\right)} = \frac{e W}{m c \gamma} \frac{\tilde{A}}{4} \frac{\omega_c^2}{v_{pl}^2}\left(\frac{W}{n\pi}\right)^2 \cos\varphi \sin\theta \cos\theta.
\end{equation}
Thus, the photocurrent has a minimum in dependence on the magnetic field if $\cos 2(\varphi - \varphi_0) > 0$. The photovoltage is a monotonously increasing function of the magnetic field.

\section{Duscussion and Conlusion}

In our work we make some approximation and now we want to discuss application conditions. All these approximation is physically intuitive but they have to be talked over separately. First, we consider the fully screened limit when the distance $d$ is much smaller than the characteristic plasmon wavelength or, more precisely, it is formulated as $d \ll W / n\pi$. In fact, this condition determines the limiting number of the plasma mode in this approximation. Second, we refuse to summarize of the entire series \eqref{eq:series} and approximate the linear velocity with only one term describing the $n$-th plasma resonance. It is applicable while the frequency difference of two nearest plasma resonances much more than the summation of their halfwidthes. Taking into account that the number $n$ is only odd we can write out this condition as $\omega_{n+2} - \omega_{n} \gtrsim \Delta\omega_{n} + \Delta\omega_{n+2}$. This is especially significant in strong magnetic field where the frequencies of all plasmons approach to the cyclotron frequency and one term is not enough. In the case we get that 
\begin{equation}
    \omega_c \lesssim \frac{v_{pl}^2}{\gamma}\frac{(n + 1) \pi^2}{W^2}.
    \label{eq:critical_magnetic_field}
\end{equation}
Third, we considered the quadratic response as a correction to the linear response. This sets limit on the allowable magnitude and, consequently, the average intensity of the incident wave $I_0$. For example, this limit has a following form
\begin{equation}
    I_0 \lesssim \frac{n_s m \gamma c^2}{W^2}\frac{2}{\order{A}{1}{n}}\frac{\left(1 + \frac{\omega_c^2}{\omega_n^2}\right)^2}{\cos^2\varphi \sin^2\theta \cos\theta},
\end{equation}
which can be obtained from the smallness the integral photocurrent $|J_n|$ relative to the root mean square current $n_s|e|\sqrt{\langle \bm{v}_n | \bm{v}_n \rangle}$.

\begin{figure}
	\centering
	\includegraphics[width=\linewidth]{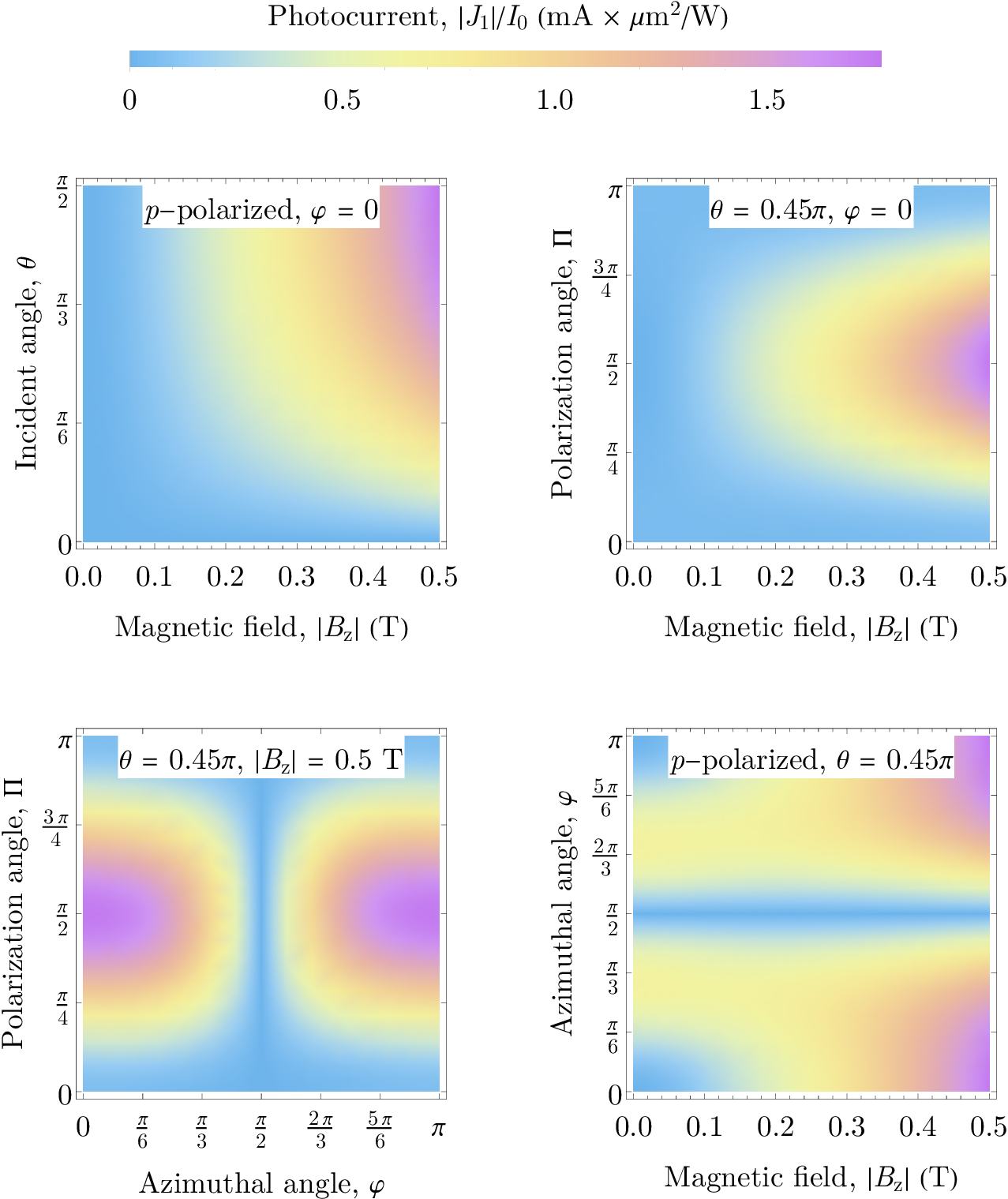}
	\caption{The absolute photocurrent $|J_1|$ divided by the average intensity of the wave $I_0$ at the first magnetoplasmon frequency. Data is presented at $n_s=3\times10^{11}$ cm$^{-2}$, $\gamma = 10^{10}$ c$^{-1}$, $W=10$ $\mu$m, $d=500$ nm, $\varepsilon = 12.7$.}
	\label{fig:3}
\end{figure}

At the end we make some quantitative estimates. For simplicity, we discuss only the excited plasma mode with $n=1$ below. As a 2D ES, we consider quantum wells based on GaAs/AlGaAs which can be potentially applied in detection the subterahertz radiation at the ``helium'' temperature $4.2$ K \cite{Muravev2012}. The effective mass of the electrons are equal to 0.068 of the mass of the free electron. The permittivity $\varepsilon$ of the AlGaAs substrate is 12.7. For example, we consider the following parameters of the 2D ES which is achievable in practice: $n_s=3\times10^{11}$ cm$^{-2}$, $\gamma = 10^{10}$ c$^{-1}$, $d=500$ nm. In this structure the plasmon velocity $2.4 \times 10^8$ cm/c. To satisfy the fully screened limit, let the strip width be 10 $\mu$m. In the case the plasmon is high quality because its frequency even in the absence of magnetic field (120 GHz) significantly exceeds $\gamma/2\pi$. In Fig.~\ref{fig:3} we present dependence of the current on the direction of the wave vector, polarization of the wave and magnetic field in different cases. For example, the p-polarized wave with $\bm{k}_{||}$ along only the strip induces the maximum photocurrent if falling is almost tangential. To satisfy the condition \eqref{eq:r_coefficients_cond} the incident angle can be selected as $0.45\pi$. To estimate the absolute values of the photoresponse, we assume that the radiant power is 10 mW, and the strip length coincides with half the wavelength along the strip, i.e., 316 $\mu$m. In this case, although the photocurrent is 5.6 nA, the photovoltage equals 5.8 $\mu$V and is a detectable value.

\section{Acknowledgments}

This work was supported by the Russian Science Foundation (Project No. 25-22-00450). We are grateful to S.A. Andreeva and I.V. Andreev for valuable discussions.

\section{Data availability}
The data that support the findings of this article are openly available \footnote{The data for ``Magnetoplasmon-Mediated Resonant Photogalvanic Effect in a Gated Strip of 2D Electrons'', \url{https://github.com/danilrodionov/FSL-photoresponse-strip} (2025)}.

\appendix

\section{\label{app}The external electromagnetic field}

Here we get the expression for the external electromagnetic radiation including the incident monochromatic wave \eqref{eq:E_inc} and its reflection from the dielectric and metal. According to Maxwell's equations, we have:
\begin{gather}
    \text{rot}\,\bm{\mathcal{E}}^{ext}(\bm{r},z,t) = \frac{\partial}{\partial t}\bm{\mathcal{B}}^{ext}(\bm{r},z,t),
    \\
    \text{rot}\,\bm{\mathcal{B}}^{ext}(\bm{r},z,t) = -\varepsilon(z)\frac{\partial}{\partial t}\bm{\mathcal{E}}^{ext}(\bm{r},z,t).
    \nonumber
\end{gather}
They can be reduced to the equation for only the electric field:
\begin{equation}
    \left[\text{grad}\,\text{div} - \Delta +\varepsilon(z)\frac{\partial^2}{\partial t^2}\right]\mathcal{E}^{ext}(\bm{r},z,t) = 0.
    \label{eq:Helmholtz}
\end{equation}
Since the system is heterogeneous only along the $z$-direction we can find the field as
\begin{equation}
    \bm{\mathcal{E}}^{ext}(\bm{r},z,t) = \frac{1}{2}\bm{\mathcal{E}}^{ext}(z) e^{i\bm{k}_{||}\bm{r} - i\omega t} + c.c.
    \nonumber
\end{equation}
where $\bm{\mathcal{E}}^{ext}(z)$ is a complex function and $\omega$ is the frequency. Further, we get the expression for $\bm{\mathcal{E}}^{ext}$. In the case, the equation can be written in the following form
\begin{multline}
    \begin{pmatrix}
        -\bm{k}_{||}\otimes\bm{k}_{||} - \left(-k_{||}^2 + \frac{\partial^2}{\partial z^2} + \varepsilon(z)\frac{\omega^2}{c^2}\right)\hat I & i\bm{k}_{||}\frac{\partial}{\partial z}
        \\
        i\bm{k}_{||}^T\frac{\partial}{\partial z} & k_{||}^2 - \varepsilon(z)\frac{\omega^2}{c^2}
    \end{pmatrix}\cdot
    \\
    \begin{pmatrix}
        \bm{\mathcal{E}}_{||}^{ext}(z)
        \\
        \mathcal{E}_{z}^{ext}(z)
    \end{pmatrix}
    = 0,
    \nonumber
\end{multline}
where $\otimes$ is a tensor product. The third string helps to get relation between the $z$-components and the others:
\begin{equation}
    \mathcal{E}_z^{ext}  = i\frac{\bm{k}_{||} \cdot \frac{\partial}{\partial z}\bm{\mathcal{E}}_{||}^{ext}(z)}{\varepsilon(z)\frac{\omega^2}{c^2} - k_{||}^2}.
    \nonumber
\end{equation}
It allows to derive the equation for the tangential component of the electric field:
\begin{multline}
    \Bigg[\frac{\partial}{\partial z}\left(\hat I - \frac{\bm{k}_{||}\otimes\bm{k}_{||}}{k_{||}^2 - \varepsilon(z)\frac{\omega^2}{c^2}}\right)\frac{\partial}{\partial z} - \left(k_{||}^2 - \varepsilon(z)\frac{\omega^2}{c^2}\right)\hat I + 
    \\
    \bm{k}_{||}\otimes\bm{k}_{||}\Bigg]
    \bm{\mathcal{E}}_{||}^{ext}(z) = 0.
    \nonumber
\end{multline}
Non diagonal elements of the obtained differential operator contains only in the matrix $\bm{k}_{||} \otimes \bm{k}_{||}$. Consequently, it is sufficient to diagonalize it with the orthogonal transformation $S$:
\begin{equation}
    S = \frac{1}{k_{||}}
    \begin{pmatrix}
        k_x & k_y\\
        -k_y & k_x
    \end{pmatrix}
    ,
    \quad
    S^{-1} = S^{T},
    \nonumber
\end{equation}
to make the entire differential operator diagonal. For the transformed electric field we have the following equation:
\begin{multline}
    \begin{pmatrix}
        \frac{\partial}{\partial z}\frac{\varepsilon(z)\frac{\omega^2}{c^2}}{\varepsilon(z)\frac{\omega^2}{c^2} - k_{||}^2}\frac{\partial}{\partial z} + \varepsilon(z)\frac{\omega^2}{c^2} & 0
        \\
        0 & \frac{\partial^2}{\partial z^2} + \varepsilon(z)\frac{\omega^2}{c^2} - k_{||}^2
    \end{pmatrix}
    \cdot
    \\
    S\bm{\mathcal{E}}_{||}^{ext}(z) = 0.
    \label{eq:SE}
\end{multline}
The first and second components of the field $S\bm{\mathcal{E}}_{||}^{ext}(z)$ are parallel and perpendicular to the tangential projection of the wavevector $\bm{k}_{||}$, respectively. To find the solution we consider separately two different spaces $z > 0$ and $0 > z \ge -d$ where the system has a simple form:
\begin{align}
    z > 0:& \quad \left(\frac{\partial^2}{\partial z^2} + k_z^2\right)S\bm{\mathcal{E}}_{||}^{ext}(z) = 0,
    \\
    0 > z \ge -d:& \quad \left(\frac{\partial^2}{\partial z^2} + K_z^2\right)S\bm{\mathcal{E}}_{||}^{ext}(z) = 0,
    \nonumber
\end{align}
Here we introduce $K_z$ as the $z$-component of the wavevector of plane waves in the dielectric substrate:
\begin{equation}
    K_z = \sqrt{\varepsilon\frac{\omega^2}{c^2} - k_{||}^2} = \frac{\omega}{c}\sqrt{\varepsilon - \sin^2\theta}.
    \nonumber
\end{equation}
In the vacuum we find solution which corresponds to the superposition of the incident and reflected waves. In the metal plane the tangential electric field has to equal zero. Taking it into account we can write the solution in the following form:
\begin{align}
    z > 0:& \quad S\bm{\mathcal{E}}_{||}^{ext}(z) = S\bm{\mathcal{E}}_{||}^{inc} e^{i k_z z} + S\bm{\mathcal{E}}_{||}^{ref} e^{-i k_z z},
    \nonumber
    \\
    0 > z \ge -d:& \quad S\bm{\mathcal{E}}_{||}^{ext}(z) = S\bm{\mathcal{E}}_{||}^{sub}\sin K_z(z + d),
\end{align}
where $\bm{\mathcal{E}}^{inc}_{||}$, $\bm{\mathcal{E}}^{ref}_{||}$ and $\bm{\mathcal{E}}^{sub}_{||}$ are amplitudes at that the last two ones are to be found. To derive them we consider the tangential electric field in the vicinity of the plane $z = 0$ where it has to satisfy two conditions. Firstly, it is continuous according to Maxwell's equations. Secondly, its derivative has a jump which can be found by integrating Eq.~\eqref{eq:SE} between $z = -0$ and $z = +0$:
\begin{multline}
    \begin{pmatrix}
        \frac{\omega^2}{k_z^2c^2} & 0
        \\
        0 & 1
    \end{pmatrix}
    \frac{\partial}{\partial z}S\bm{\mathcal{E}}_{||}^{ext}(z)\Big|_{z = +0}
    - 
    \\
    \begin{pmatrix}
        \frac{\varepsilon\omega^2}{K_z^2c^2} & 0
        \\
        0 & 1
    \end{pmatrix}
    \frac{\partial}{\partial z}S\bm{\mathcal{E}}_{||}^{ext}(z)\Big|_{z = -0} = 0
    \nonumber
\end{multline}
It allows to obtain the amplitude 
\begin{gather}
    S\bm{\mathcal{E}}_{||}^{ref} = 
    \begin{pmatrix}
        -r_p & 0
        \\
        0 & r_s
    \end{pmatrix}
    S\bm{\mathcal{E}}_{||}^{inc}
    \\
    S\bm{\mathcal{E}}_{||}^{sub} = \frac{1}{\sin K_z d}
    \begin{pmatrix}
        1 - r_p & 0
        \\
        0 & 1 + r_s
    \end{pmatrix}
    S\bm{\mathcal{E}}_{||}^{inc},
\end{gather}
which we write using the amplitude reflection coefficients $r_s$ and $r_p$
\begin{equation}
    r_p = -\frac{i + \varepsilon \frac{k_z}{K_z}\cot K_z d}{i - \varepsilon \frac{k_z}{K_z}\cot K_z d}, 
    \quad
    r_s = \frac{i \frac{k_z}{K_z} + \cot K_z d}{i \frac{k_z}{K_z} - \cot K_z d}.
\end{equation}
Expanding $r_s$ and $r_p$ up to the second order by the parameter $\omega d/c$ we have
\begin{gather}
    r_p = 1 + 2 i \frac{K_z^2}{k_z}d - 2\frac{K_z^4}{k_z^2}d^2 + ...,
    \\
    r_p = -1 - 2 i k_z d + 2 k_z^2 d^2 + ...
\end{gather}
The smallness of the terms $\propto d^2$ relative to the terms $\propto d$ help the acceptable incident angles. As result, in the fully screened limit we go to Eq.~\eqref{eq:r_coefficients} with restriction Eq.~\eqref{eq:r_coefficients_cond}.

\bibliography{main}

\end{document}